\documentclass[twocolumn,showpacs,amsmath,amssymb,prl,superscriptaddress,floatfix]{revtex4}
\usepackage{graphicx}
\usepackage{dcolumn}
\usepackage{bm}
\usepackage{amsfonts}
\usepackage{amsmath}
\usepackage{amssymb}

\newcommand{\fmn}[2]{\mbox{${\textstyle \frac{#1}{#2}}$}}
\newcommand{\half}{\mbox{${\textstyle \frac{1}{2}}$}}           
\begin{document}
\title{Observation of an ABC effect in proton-proton collisions}
\author{S.~Dymov}\email{s.dymov@fz-juelich.de}
\affiliation{Laboratory of Nuclear Problems, Joint Institute for
Nuclear Research, 141980 Dubna, Russia}%
\affiliation{Physikalisches Institut II, Universit{\"a}t
Erlangen-N{\"u}rnberg, 91058 Erlangen, Germany }
\author{M.~Hartmann}
\affiliation{Institut f\"ur Kernphysik, Forschungszentrum J\"ulich
GmbH, 52425 J\"ulich, Germany} %
\affiliation{J\"ulich Centre for Hadron Physics, 52425 J\"ulich,
Germany}
\author{A.~Kacharava}
\affiliation{Institut f\"ur Kernphysik, Forschungszentrum
J\"ulich GmbH, 52425 J\"ulich, Germany}%
\affiliation{J\"ulich Centre for Hadron Physics, 52425 J\"ulich,
Germany}
\author{A.~Khoukaz}
\affiliation{Institut f\"ur Kernphysik, Universit\"at M\"unster,
48149 M\"unster, Germany}
\author{V.~Komarov}
\affiliation{Laboratory of Nuclear Problems, Joint Institute for
Nuclear Research, 141980 Dubna, Russia}
\author{P.~Kulessa}
\affiliation{H.~Niewodnicza\'{n}ski Institute of Nuclear Physics PAN,
31342 Krak\'{o}w, Poland}
\author{A.~Kulikov}
\affiliation{Laboratory of Nuclear Problems, Joint Institute for
Nuclear Research, 141980 Dubna, Russia}
\author{V.~Kurbatov}
\affiliation{Laboratory of Nuclear Problems, Joint Institute for
Nuclear Research, 141980 Dubna, Russia}
\author{G.~Macharashvili}
\affiliation{High Energy Physics Institute, Tbilisi State University,
0186 Tbilisi, Georgia} \affiliation{Laboratory of Nuclear Problems,
Joint Institute for Nuclear Research, 141980 Dubna, Russia}
\author{S.~Merzliakov}
\affiliation{Laboratory of Nuclear Problems, Joint Institute for
Nuclear Research, 141980 Dubna, Russia}%
\affiliation{Institut f\"ur Kernphysik, Forschungszentrum J\"ulich
GmbH, 52425 J\"ulich, Germany}%
\affiliation{J\"ulich Centre for Hadron Physics, 52425 J\"ulich,
Germany}
\author{M.~Mielke}
\affiliation{Institut f\"ur Kernphysik, Universit\"at M\"unster,
48149 M\"unster, Germany}
\author{S.~Mikirtychiants}
\affiliation{Institut f\"ur Kernphysik, Forschungszentrum J\"ulich
GmbH, 52425 J\"ulich, Germany} %
\affiliation{J\"ulich Centre for Hadron Physics, 52425 J\"ulich,
Germany} \affiliation{High Energy Physics Department, Petersburg
Nuclear Physics Institute, 188350 Gatchina, Russia}
\author{M.~Nekipelov}
\affiliation{Institut f\"ur Kernphysik, Forschungszentrum J\"ulich
GmbH, 52425 J\"ulich, Germany} %
\affiliation{J\"ulich Centre for Hadron Physics, 52425 J\"ulich,
Germany}
\author{M.~Nioradze}
\affiliation{High Energy Physics Institute, Tbilisi State University,
0186 Tbilisi, Georgia}
\author{H.~Ohm}\affiliation{Institut f\"ur Kernphysik, Forschungszentrum
J\"ulich GmbH, 52425 J\"ulich, Germany} %
\affiliation{J\"ulich Centre for Hadron Physics, 52425 J\"ulich,
Germany}
\author{F.~Rathmann}
\affiliation{Institut f\"ur Kernphysik, Forschungszentrum J\"ulich
GmbH, 52425 J\"ulich, Germany}%
\affiliation{J\"ulich Centre for Hadron Physics, 52425 J\"ulich,
Germany}
\author{H.~Str\"oher}
\affiliation{Institut f\"ur Kernphysik, Forschungszentrum
J\"ulich GmbH, 52425 J\"ulich, Germany}%
\affiliation{J\"ulich Centre for Hadron Physics, 52425 J\"ulich,
Germany}
\author{D.~Tsirkov}
\affiliation{Laboratory of Nuclear Problems, Joint Institute for
Nuclear Research, 141980 Dubna, Russia}
\affiliation{Institut f\"ur Kernphysik, Forschungszentrum
J\"ulich GmbH, 52425 J\"ulich, Germany}%
\affiliation{J\"ulich Centre for Hadron Physics, 52425 J\"ulich,
Germany}
\author{Yu.~Uzikov}
\affiliation{Laboratory of Nuclear Problems, Joint Institute for
Nuclear Research, 141980 Dubna, Russia}
\author{Yu.~Valdau}
\affiliation{Institut f\"ur Kernphysik, Forschungszentrum
J\"ulich GmbH, 52425 J\"ulich, Germany}%
\affiliation{J\"ulich Centre for Hadron Physics, 52425 J\"ulich,
Germany}
\affiliation{High Energy Physics Department, Petersburg Nuclear
Physics Institute, 188350 Gatchina, Russia}
\author{C.~Wilkin}
\affiliation{Physics and Astronomy Department, UCL, London, WC1E 6BT,
UK}
\author{S.~Yaschenko}
\affiliation{Laboratory of Nuclear Problems, Joint Institute for
Nuclear Research, 141980 Dubna, Russia}%
\affiliation{Physikalisches Institut II, Universit{\"a}t
Erlangen-N{\"u}rnberg, 91058 Erlangen, Germany}
\author{B.~Zalikhanov}
\affiliation{Laboratory of Nuclear Problems, Joint Institute for
Nuclear Research, 141980 Dubna, Russia}
\date{\today}

\begin{abstract}
The cross section for inclusive multipion production in the $pp\to
ppX$ reaction was measured at COSY-ANKE at four beam energies, 0.8,
1.1, 1.4, and 2.0~GeV, for low excitation energy in the final $pp$
system, such that the diproton quasi-particle is in the $^{1\!}S_0$
state. At the three higher energies the missing mass $M_X$ spectra
show a strong enhancement at low $M_X$, corresponding to an ABC
effect that moves steadily to larger values as the energy is
increased. Despite the missing-mass structure looking very different
at 0.8~GeV, the variation with $M_X$ and beam energy are consistent
with two-pion production being mediated through the excitation of two
$\Delta(1232)$ isobars, coupled to $S$-- and $D$-- states of the
initial $pp$ system.
\end{abstract}

 \maketitle

The ABC effect is a sharp enhancement of the two-pion invariant mass
spectrum near its threshold that has puzzled hadron physicists for
almost fifty years. First observed in a $pd\to \,^{3}\textrm{He}X$
missing-mass experiment~\cite{Abashian}, and subsequently confirmed
using a deuteron beam~\cite{Banaigs1}, it was seen most spectacularly
in the $dd\to \,^{4}\textrm{He}X$ reaction~\cite{Banaigs2,Wurzinger}.
The ABC typically manifests itself as a peak at an invariant mass
$M_X\approx 310-330$~MeV/$c^2$ and it was first even speculated that
it might be a scalar meson with isospin $I=0$. However, the fact that
the mass and width varied with experimental
conditions~\cite{Banaigs1} suggests that the ABC is a kinematic
enhancement. This is consistent with the smooth behavior of the
isoscalar $\pi\pi$ phase shifts at low energies. To understand the
nature of the ABC it is necessary to investigate the effect in
simpler systems.

The ABC showed up very clearly in the zero-degree momentum spectrum
of the deuteron from $np\to dX$~\cite{Plouin}. Here it leads to two
peaks, \emph{i.e.}\ forward and backward production in the c.m.\
system, though the momentum spread of the neutron beam degraded the
missing-mass resolution. This has recently been overcome by measuring
the quasi-free $pd \to d\pi^0\pi^0p_{\,\text{sp}}$ reaction
semi-exclusively at beam energies of $T_p=1.03$ and 1.35~GeV, with
just the spectator proton $p_{\,\text{sp}}$ escaping
detection~\cite{WASA}. The only published study of the ABC effect in
proton-proton collisions was carried out at $T_p=1.52$ and
1.81~GeV~\cite{Yonnet}, though this concentrated on possible
substructure.

Studies of the energy dependence of the $pd\to \,^{3}\textrm{He}X$
and $dd\to \,^{4}\textrm{He}X$ cross sections showed that the ABC
effect is most prominent for energies where the maximum missing mass
is about 600~MeV/$c^2$~\cite{Banaigs2}. Since this corresponds to the
mass difference between two $\Delta(1232)$ isobars and two nucleons,
it is tempting to suppose that the two-pion production is mediated by
the excitation and decay of two separate $\Delta$ isobars. Such a
model with $\pi$~\cite{Malcolm} and then $\pi+\rho$
exchange~\cite{Osterfeld} had semi-quantitative success in describing
the existing $np\to dX$ data. The dominant contribution to the cross
section arises when the two $\Delta$ are in a relative $S$-wave. For
the production of an $I_{\pi\pi}=0$ pion pair, isospin conservation
requires $I_{\Delta\Delta}=0$. It follows from the generalized Pauli
principle that the two isobars must then couple to a total spin of
$S_{\Delta\Delta}=1$ or 3. In contrast, for $pp\to \Delta\Delta\to
pp\pi\pi$, $I_{\Delta\Delta}=1$ so that $S_{\Delta\Delta}=0$ or 2,
and an $s$--wave pion pair can have $I_{\pi\pi}=0$ or 2. The
investigation of the ABC phenomenon in $np$ and $pp$ collisions is
therefore largely complementary.

The aim of the present experiment is the measurement of the $pp\to
\{pp\}_{\!s}X$ reaction when the excitation energy in the final
diproton system is low, $E_{pp}<3$~MeV. This quasi-particle, denoted
by $\{pp\}_{\!s}$, will be almost exclusively in the $^{1\!}S_0$
state. Under these conditions, the kinematics are similar to those of
$pn \to dX$, which facilitates a comparison of the two reactions. Of
the four proton beam energies used, $T_{p}=0.8$, $1.1$, $1.4$, and
$2.0$~GeV, the lowest is well below the nominal $\Delta\Delta$
threshold, and by 2~GeV much of the $\Delta$ strength has passed.

The measurement of single and multipion production in $pp$ collisions
was performed using the ANKE magnetic spectrometer~\cite{Barsov},
positioned at an internal target station of the COSY synchrotron
storage ring of the Forschungszentrum J\"{u}lich. The single pion
results have already been reported~\cite{Dymov1}. The proton beam was
directed at a hydrogen cluster-jet target with an areal density of
$2\times10^{14}$~atoms/cm$^{2}$. The resulting positively charged
particles were recorded in the ANKE forward detector system,
consisting of three MWPCs and a two-layer scintillation hodoscope,
the whole giving a horizontal acceptance of $12^{\circ}$ and
$\pm3.5^{\circ}$ vertically. The particle momenta, deduced by tracing
through the analyzing magnetic field, had typical uncertainties of
$\approx 1\%$. The emerging proton pairs were identified by
evaluating the difference $\Delta t$ between the times of flight
measured with the hodoscope and those calculated from the particle
momenta, assuming the particles to be protons~\cite{Dymov1}.

\begin{figure}[htb]
\vspace*{1mm}
\centering
\includegraphics*[width=7cm]{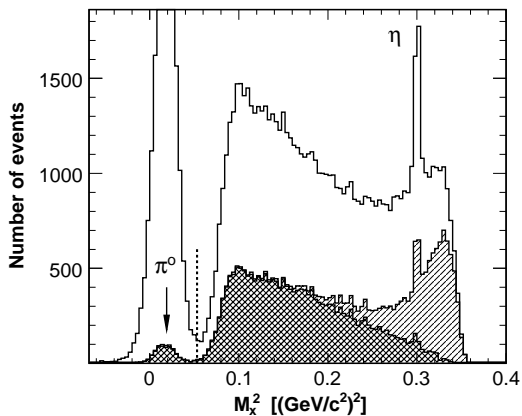}
\caption{Distribution in the square of the missing mass of the $pp\to
ppX$ reaction at $1.4$~GeV. Imposing a $\Delta t$ cut to select the
two protons gives the open histogram. The lightly shaded plot
corresponds to the additional requirement that $E_{pp}<3$~MeV and the
heavy shading reflects the further $\cos\vartheta_{pp}>0.95$ cut to
match the overall ANKE acceptance. The positions of the $\pi^0$ and
$\eta$ peak are indicated, as is the two-pion threshold (dotted
line).} \label{f:Mx2.14}
\end{figure}

Figure~\ref{f:Mx2.14} shows the spectrum in the square of the missing
mass of the $pp\to ppX$ reaction at 1.4~GeV. Selecting $pp$ events on
the basis of the $\Delta t$ cut eliminates almost all $d\pi^{+}$
pairs, leaving a background of less than 0.1\%. To ensure that the
two protons are dominantly in the $^{1\!}S_{0}$ state, a further cut
on the $pp$ excitation energy, $E_{pp} < 3$~MeV, was imposed. After
weighting by detection efficiency, such pairs are distributed
isotropically in their rest frame, as expected for an $S$-wave. The
$E_{pp}$ distributions also follow closely the expected $S$-wave
final-state-interaction behavior. Since for low missing masses the
c.m.\ angle of the diproton, $\vartheta_{pp}$, is strongly limited by
the ANKE acceptance, we imposed the extra cut
$\cos\vartheta_{pp}>0.95$ on the whole spectrum. Apart from peaks
corresponding to the production of single $\pi^0$ and $\eta$ mesons,
the distribution shows a rapid rise from the two-pion threshold with
a gentle decrease towards the upper kinematic limit.

\begin{figure}[hptb]
\vspace*{1mm}
\centering
\includegraphics*[width=7cm]{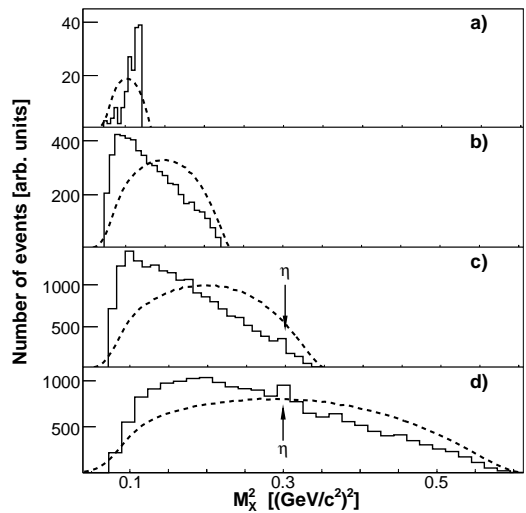}
\caption{Distribution in missing-mass squared for the $pp\to
\{pp\}_{\!s}X$ reaction for $E_{pp}<3$~MeV and
$\cos\vartheta_{pp}>0.95$ at a) 0.8, b) 1.1, c) 1.4, and d) 2.0~GeV.
The $\eta$ signal is seen at the expected position for the two higher
energies. The curves represent normalized simulations within a
phase-space model. \label{raw}}
\end{figure}

The development of the counting rate distribution with energy is
displayed in Fig.~\ref{raw} for the multipion region. Also shown
there are the results of a phase-space Monte-Carlo simulation of the
detector and analysis code, including effects of momentum resolution,
and applying the same cuts as for the experiment. In no case does
this approach reproduce the data.

\begin{figure}[hptb]
\vspace*{1mm}
\centering
\includegraphics*[height=8cm,width=0.95\columnwidth]{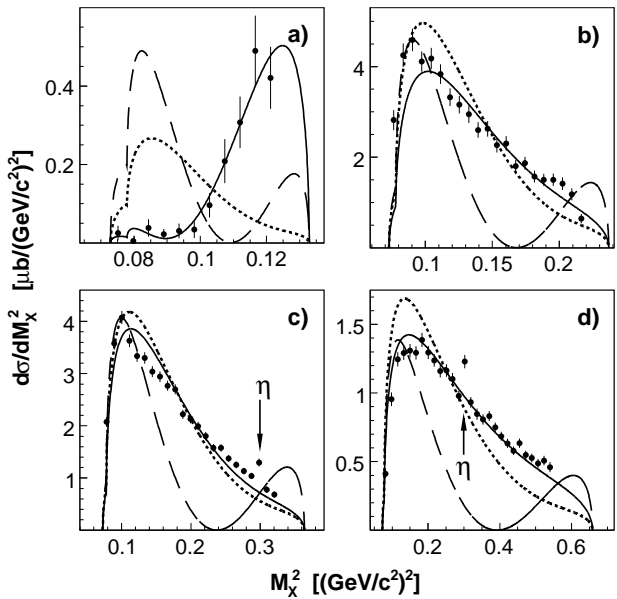}
\caption{The $pp\to ppX$ differential cross section with statistical
errors as a function of the square of the missing mass at a) 0.8, b)
1.1, c) 1.4, and d) 2.0~GeV for $E_{pp}<3$~MeV and
$\cos\vartheta_{pp}>0.95$. The $\eta$ peaks are indicated. Events
were simulated using Eq.~(\ref{final}) with $A_D=0$ (long dashes),
with $A_S=0$ (short-dashed), and with the best fit of
Table~\ref{tab1} (solid line).} \label{dcex}
\end{figure}

A simulation was used to deduce the differential cross sections of
Fig.~\ref{dcex} from the counting rate data, taking as input the
double-$\Delta$ model to be discussed with Eq.~(\ref{final}). The
simulation also defines the momentum reconstruction uncertainties and
this leads to a $\pi^{0}$ peak whose width is compatible with that
shown in Fig.~\ref{f:Mx2.14}. The requisite luminosity was derived
from the simultaneous measurement of $pp$ elastic
scattering~\cite{Dymov1}.

As shown in Fig.~\ref{dcex}, the strength at 0.8~GeV is pushed
towards the maximum missing mass, while at the higher three energies
there is a peak of the ABC type at low $M_X$ and no sign of any
enhancement at large masses. Above its threshold there are also
contributions from $\eta$ production, with cross sections integrated
over our cuts of $\sigma_{\eta} = 4.3\pm 0.8$ at 1.4~GeV and
$4.5\pm0.9$~nb at 2.0~GeV. The ABC position moves steadily to the
right and gets broader as $T_p$ increases above 1.1~GeV, a
confirmation that the phenomenon is primarily a kinematic effect. The
peak seems to be rather clearer than that observed at Saclay at 1.5
and 1.8~GeV~\cite{Yonnet}. Although a previous exclusive measurement
of $pp\to pp\pi^+\pi^-$ at 0.8~GeV~\cite{TOF} did not study in detail
our kinematic configuration, the two data sets are consistent in the
region of overlap.

The total cross section for the production of three pions in $pp$
collisions at 1.35~GeV is about 0.6\% of that for two
pions~\cite{Pauly} and so it is only at 2~GeV that there might be any
significant contribution to the data from the $pp\to pp\pi\pi\pi$
reaction. The modelling of the results must therefore be based upon
the assumption that two-pion production dominates.

A $\Delta\Delta$ model applied to two-meson production in the $pn\to
dX$ or $dd\to\alpha X$ reaction allows the momentum transfer to be
shared between the nucleons, facilitating the formation of the
observed deuteron or $\alpha$-particle. The strong peaking of the
$\Delta$ decay in the forward and backward directions also enhances
this sticking factor. It is therefore particularly suited to our
conditions but might not be appropriate for other kinematic domains.
A full estimation of two-pion production in such a model depends
sensitively upon the exchange mesons and form factors \emph{etc}.
Nevertheless, the general structure can be written down if we assume
that the intermediate $\Delta$ are in a relative $S$-wave so that the
initial protons are in the $^{1\!}S_0$ or $^{1\!}D_2$ wave. Since the
$p$-wave nature of each $\Delta$ decay necessarily involves a
momentum factor, the obvious \emph{ansatz} for the
$pp\to\{pp\}_{\!s}\pi\pi$ matrix element is
\begin{equation}
\label{eq1} \mathcal{M}=A_S(\vec{k}'_1\cdot\vec{k}'_2)
+\fmn{1}{2}A_D\left(3(\hat{p}\cdot\vec{k}'_1)(\hat{p}\cdot\vec{k}'_2)
-\vec{k}'_1\cdot\vec{k}'_2\right),
\end{equation}
where $\vec{p}$ is the c.m.\ proton beam momentum and $\vec{k}'_i$ is
the relative momentum of the $i$'th pion with respect to one of the
recoiling protons. The $S$ and $D$--wave amplitudes $A_S$ and $A_D$
are scalar functions that may depend strongly upon the kinematic
variables because of the excitation of the two $\Delta$, though we
shall neglect any variation with $M_X$. In terms of the relative
$\pi\pi$ momentum, $2\vec{q}$, and that of the diproton, $-\vec{k}$,
the pion c.m.\ momenta are $\vec{k}_{1,2}=\half\vec{k}\pm\vec{q}$,
and the matrix element becomes
\begin{eqnarray}
\nonumber
\mathcal{M}&=&A_S(\alpha^2k^2-\beta^2q^2)\\
\label{eq2}
&&\hspace{-2.5cm}+\fmn{1}{2}A_D\left(3(\alpha^2k_z^2-\beta^2q_z^2)
- (\alpha^2k^2-\beta^2q^2)\right),
\end{eqnarray}
where the $z$-direction is taken along $\vec{p}$. In this
approximation, $q^2$ and $k^2$ are linked by energy conservation. For
simplicity of presentation, a Galilean transformation is used to
evaluate the recoil factors $\alpha=\half(m+2\mu)/(m+\mu)$ and
$\beta=m/(m+\mu)$ in terms of the pion ($\mu$) and proton ($m$)
masses. Only the part of the matrix element proportional to
$3q_z^2-q^2$ corresponds to a $d$-wave in the $\pi\pi$ system and
this contributes very little to the ABC peak, which is $s$-wave in
nature.

Since the direction of the $\pi\pi$ relative momentum is not observed
in a missing-mass experiment, $|\mathcal{M}|^2$ must be averaged over
the angles of $\vec{q}$;
\begin{eqnarray}
\nonumber <|\mathcal{M}|^2>&=&\left|A_S(\alpha^2k^2-\beta^2q^2)
+\fmn{1}{2}A_D\alpha^2(3k_z^2-k^2)\right|^2\\
&&\hspace{-0.5cm}+\fmn{1}{5}|A_D|^2\beta^4q^4\,. \label{final}
\end{eqnarray}

Figure~\ref{dcex} shows the results of implementing the simple
double--$\Delta$ model of Eq.~(\ref{final}). Since the charges of the
pions were not measured, we assumed an $I_{\pi\pi}=0$ ratio of
$\pi^+\pi^-$:$\,\pi^0\pi^0 = 2:1$, with the corresponding masses
being used in the two phase spaces. If there were only $S$-wave
production, $<|\mathcal{M}|^2>$ would be maximal at $q^2=0$ (the ABC
bump) and at $k^2=0$ (the maximum allowed missing mass), with a deep
valley between these two structures. Such behavior is seen for
$dd\to\,^{4}\textrm{He}X$~\cite{Banaigs2,Wurzinger}. The $A_D=0$ case
is illustrated in Fig.~\ref{dcex}, as is that of a pure $D$-wave,
which gives a broad maximum at low $\pi\pi$ invariant masses and no
sign of any high mass bump.

The general shapes of the spectra at the three higher energies are
qualitatively reproduced by the pure $D$-wave model. However, the
peaking towards maximum $M_X$ in the 0.8~GeV data shows the necessity
for a substantial $S$-wave contribution at this energy. To
investigate this, the uncorrected data of Fig.~\ref{raw} were fitted
to the Monte-Carlo simulations using as input Eq.~(\ref{final}) with
free values of $A_S$ and $A_D$. The values of these parameters that
best reproduce the spectra are given in Table~\ref{tab1} along with
the resulting integrated cross sections. Only at 0.8~GeV is $A_S/A_D$
required to be large. The description of the data shown in
Fig.~\ref{dcex} is much improved, especially at 0.8~GeV, but the
model does not reproduce the sharpness of the ABC peaks seen at 1.1
and 1.4~GeV. The values of $A_D$ fall steadily with energy but, to
understand the significance of this, would require a full dynamical
model.

\begin{table}[htb]
\caption{Cross section $\sigma_X$ for the $pp\to \{pp\}_{\!s}X$
reaction integrated over $\cos\vartheta_{pp}>0.95$, $E_{pp}<3$ MeV
and over $M_X$ up to the kinematical limit. The experimental
acceptance is limited to less than $M_X^\textrm{max}$, and $\sigma_X$
is obtained by correcting the measured cross-section by an amount
$\xi$, as suggested by fits of the $\Delta\Delta$ model to the
differential data. In addition to the given statistical errors, there
are systematic uncertainties of 7\%, coming principally from the
luminosity and acceptance evaluations~\cite{Dymov1}. The parameters
$A_S$ and $A_D$ are determined by fitting the spectra away from
regions of rapid variation, where resolution questions arise, and at
too high masses to avoid the tail from $\rho$ production. The results
for $A_D$ are normalized at 1.1~GeV.}
\begin{ruledtabular}
\begin{tabular}{cccccc}
$T_p$ &$M_X^{\textrm{max}}$     & $\sigma_X$   & $\xi$ & $A_S/A_D$   & $A_D$\\
GeV &$\textrm{MeV/c}^2$& nb         & $\%$ &             &      \\
\hline
0.8   & 351 & $\phantom{1}12\pm 2$ & 34 & $-1.23\pm0.10$ & $1.09\pm0.05$\\
1.1   & 468 & $389\pm 6$& 2.3  & $-0.20\pm0.03$ & $1.00$\\
1.4   & 571 & $563\pm 6$& 2.4  & $-0.10\pm0.03$ & $0.48\pm0.01$\\
2.0   & 741 & $456\pm 5$& 5.8  & $-0.23\pm0.02$ & $0.17\pm0.01$\\
\end{tabular}
\end{ruledtabular}
\label{tab1}
\end{table}

The special quasi-particle kinematics studied here are very similar
to those of $pn \to d\pi^0\pi^0$ and so a comparison with the
CELSIUS-WASA data~\cite{WASA} is instructive because of the different
spin-isospin selection rules. The measurement of the Fermi momentum
when using a deuterium target allowed the determination of cross
sections as a function of the laboratory kinetic energy $T_p$. The
results suggested a resonance-like structure with a mass of
2.39~GeV/$c^2$ and width of $\Gamma \approx 90$~MeV/$c^2$
($T_p\approx 1.17$~GeV, $\Gamma_T = 0.23$~GeV). Though our energy
intervals are rather wide, no such behavior is obvious in the values
given in Table~\ref{tab1}, which would suggest a strong isospin
dependence of the production mechanism.

Within the $S$--wave $\Delta\Delta$ approach, the matrix element of
Eq.~(\ref{eq1}) allows for both $A_S$ and $A_D$ terms. If two-pion
production were driven instead by the excitation of $S$--wave
$N^*(1440)N$ pairs, where $N^*$ is the Roper resonance, an $A_S$ term
is generated through the double $p$-wave cascade decay $N^*(1440)\to
\Delta\pi\to p\pi\pi$. However, this model requires that $A_D=0$.
Contributions from both $N^*N$ and $\Delta\Delta$ might therefore
provide an explanation for the energy dependence of $A_S/A_D$ shown
in Table~\ref{tab1}. It should be noted that the $\Delta\Delta$ model
would lead to an ABC effect also in $pp\to \{nn\}_{\!s}\pi^+\pi^+$,
which is forbidden in the $N^*N$ model. Earlier studies of the ABC
effect were not sensitive to any $I_{\pi\pi}=2$ component.

In summary, we have measured the differential cross section for the
$pp\to \{pp\}_{\!s}X$ reaction at four beam energies from 0.8 to
2.0~GeV under the specific kinematic conditions where the
proton-proton excitation energy is below 3~MeV and the c.m.\ angle
between the diproton momentum and the beam axis is less than
$18^{\circ}$. The form of the $E_{pp}$ spectra and isotropy of the
angular distribution in the diproton frame are consistent with the
two final protons being in the $^{1\!}S_0$ state. Strong deviations
from phase space are observed in the missing-mass variable, with a
peak in $M_X$ whose position varies with beam energy, and no broad
bump at maximum missing mass of the type observed in $dd\to\alpha
X$~\cite{Banaigs2,Wurzinger}. On the other hand, there is no sign of
the resonance-like structure in the energy dependence of the
integrated cross section that is reported for $pn\to
d\pi^0\pi^0$~\cite{WASA}.

The evidence presented here points towards to the ABC produced in
$pp$ collisions being a kinematic effect and this is backed up by our
naive phenomenological description. The isospin structure of $pp\to
\{pp\}_{\!s}X$ is much richer than in the $pn\to d X$ case and it is
quite likely that there will be significant contributions from both
$I_{\pi\pi}=0$ and 2; a kinematic enhancement does not have to have a
definite isospin. On the other hand, the spin structure is much
simpler and, after assuming that each of the two pions is emitted in
a $p$-wave, there are only two amplitudes to be considered. Except
for the 0.8~GeV data, the bulk of the $M_X$ spectra are consistent
with a dominant $D$--wave coupling to the incident protons, whereas
at the lowest energy the $S$-wave is significant. The major
discrepancies in the model are in the heights of the ABC peaks at 1.1
and 1.4~GeV, but a fully microscopic approach would be required to
assess the origin of these.

Exclusive measurements of $pp\to\{pp\}_{\!s}(\pi\pi)^0$ over a wider
range of angles would provide more stringent tests of the simple
phenomenological description and the study of
$pp\to\{nn\}_{\!s}\pi^+\pi^+$ would be particularly valuable.

We are grateful to other members of the ANKE Collaboration for their
help with this experiment and to the COSY crew for providing such
good working conditions. This work was supported in part by the BMBF
grant ANKE COSY-JINR, HGF--VIQCD, and JCHP FFE.

%
%


\begin{thebibliography}{99}
%
\bibitem{Abashian} A.~Abashian, N.~E.~Booth, and K.~M.~Crowe, Phys.\ Rev.\
  Lett.\ \textbf{5}, 258 (1960).
%
\bibitem{Banaigs1} J.~Banaigs \emph{et al.}, Nucl.\ Phys.\ \textbf{B67}, 1
  (1973).
%
\bibitem{Banaigs2} J.~Banaigs \emph{et al.}, Nucl.\ Phys.\ \textbf{B105}, 52
  (1976).
%
\bibitem{Wurzinger} R.~Wurzinger \emph {et al.}, Phys.\ Lett.\ B \textbf{445}, 423
  (1999).
%
\bibitem{Plouin} F.~Plouin \emph{et al.}, Nucl.\ Phys.\ \textbf{A302}, 413
  (1978).
%
\bibitem{WASA} M.~Bashkanov \emph{et al.}, Phys.\ Rev.\ Lett.\ \textbf{102}, 052301
(2009).
%
\bibitem{Yonnet} J.~Yonnet \emph{et al.}, Phys.\ Rev.\ C \textbf{63}, 014001 (2000).
%
\bibitem{Malcolm} T.~Risser and M.~D.~Shuster, Phys.\ Lett.\ \textbf{43B}, 68
  (1973).
%
\bibitem{Osterfeld} C.~A.~Mosbacher and F.~Osterfeld, in \textit{Baryons '98},
  Ed.\ D.~W.~Menze and B.~Metsch (World Scientific, Singapore) p.~609.
%
\bibitem{Barsov} S.~Barsov \emph{et al.}, Nucl.\ Instrum.\ Meth.\ Phys.\ Res.,
Sect.\ A \textbf{462}, 364 (2001).
%
\bibitem{Dymov1} S.~Dymov \emph{et al.}, Phys.\ Lett.\ B \textbf{635}, 270
  (2006); V.~Kurbatov  \emph{et al.}, Phys.\ Lett.\ B \textbf{661}, 22 (2008).
%
\bibitem{Pauly} C.~Pauly \emph{et al.}, Phys.\ Lett.\ B \textbf{649}, 122 (2007).
%
\bibitem{TOF} S.~Abd El-Bary \emph{et al.}, Eur.\ Phys.\ J.\ A
\textbf{37}, 267 (2008).
%
\end{thebibliography}
\end{document}